\documentclass[11pt]{article} 
\usepackage{mystyle-new}
\usepackage{authblk}
\usepackage[T1]{fontenc}
\usepackage{epsfig,amsmath} 
\usepackage{hepnames,hepunits}
\usepackage{hepunits}
\usepackage{hyperref}
\usepackage{color}
\usepackage{graphicx}
\definecolor{red}{rgb}{1,0,0}
\def\lesssim{\ \hbox{\raise 2pt \hbox{$<$} \kern -13pt
                     \lower 3pt \hbox{$\sim$}}\ }
\def\greatersim{\ \hbox{\raise 2pt \hbox{$>$} \kern -13pt
                     \lower 3pt \hbox{$\sim$}}\ }

\def\lsim{\mathrel{\rlap{\lower4pt\hbox{\hskip1pt$\sim$}}
    \raise1pt\hbox{$<$}}}                
\def\gsim{\mathrel{\rlap{\lower4pt\hbox{\hskip1pt$\sim$}}
    \raise1pt\hbox{$>$}}}                

\def\cascade{{\scshape Cascade3}}
\def\pythia{{\scshape Pythia}}
\def\herwig{{\scshape Herwig}}

\def\mcatnlo{{MCatNLO}}

\input epsf.tex
\def\desepsf(#1 width #2){\epsfxsize=#2 \epsfbox{#1}}
\def\kt{\ensuremath{k_{\rm T}}}

\def\pt{\ensuremath{p_{\rm T}}}

\def\ptZj{\ensuremath{p_{{\rm T},Zj}}}

\def\PZ{\rm{Z}\xspace}
\def\Pp{\rm{p}\xspace}

\newcommand{\alphas}{\ensuremath{\alpha_\mathrm{s}}}

\newcommand{\asmz}{\ensuremath{\alphas(m_{\PZ})}\xspace}

\newcommand{\PBM}{PB}

\newcommand{\dphi}{{\ensuremath{\Delta\phi_{12}}}}
\newcommand{\dphiZ}{{\ensuremath{\Delta\phi_{\PZ\text{j}}}}}
\newcommand{\MCatNLO}{{\scshape MadGraph5\_aMC@NLO}}
\newcommand{\ptmax}{\ensuremath{\pt^{\text{leading}}}\xspace}

\newcommand{\Zjet}{{$\PZ+$\text{jet}}}

\newenvironment{tolerant}[1]{\par\tolerance=#1\relax}{ \par }
\usepackage{amsmath,bm}
\usepackage{lineno}

\usepackage{cite,./mcite}
\usepackage{tikz}
\usepackage[symbol]{footmisc}

\newcommand{\dglap}{Gribov:1972ri,Lipatov:1974qm,Altarelli:1977zs,Dokshitzer:1977sg}

\providecommand{\DOI}[1]{\href{http://dx.doi.org/#1}}

\begin{document}

\title{
Back-to-back azimuthal correlations in \Zjet\ events at high transverse momentum 
 in the TMD parton branching method at next-to-leading order}
\author[1,2]{H.~Yang}
\affil[1]{School of Physics, Peking University}
\affil[2]{Deutsches Elektronen-Synchrotron DESY, Germany}
\author[2]{A.~Bermudez~Martinez}
\author[2]{L.I.~Estevez~Banos}
\author[3,4,5]{F.~Hautmann}
\affil[3]{CERN, Theoretical Physics Department, Geneva}
\affil[4]{Elementary Particle Physics, University of Antwerp, Belgium}
\affil[5]{University of Oxford, UK}
\author[2]{H.~Jung}
\author[2]{M.~Mendizabal}
\author[6]{K.~Moral~Figueroa}
\affil[6]{University of Edinburgh, UK}
\author[7]{S.~Prestel}
\affil[7]{Department of Astronomy and Theoretical Physics, Lund University, Sweden} 
\author[2]{S.~Taheri~Monfared}
\author[4]{A.M.~van~Kampen}
\author[2,1]{Q.~Wang}
\author[2]{K.~Wichmann}

\begin{titlepage} 
\maketitle
\vspace*{-15cm}
\begin{flushright}
CERN-TH-2022-113\\
DESY-22-025  \\
\end{flushright}
\vspace*{+13cm}
\end{titlepage}

\begin{abstract}
Azimuthal correlations in  \Zjet\  production at large transverse momenta are 
computed by matching Parton - Branching (PB) TMD parton distributions and showers with  NLO 
calculations via \mcatnlo. The predictions  are 
compared with those for dijet production in the same kinematic range. The azimuthal correlations $\Delta\phi$ between  the \PZ\ boson and the leading jet are steeper compared to those in dijet production at transverse momenta ${\cal O}(100)$~\GeV , while they become similar for very high transverse momenta ${\cal O}(1000)$~\GeV .    
 The different patterns of  \Zjet\  and dijet azimuthal correlations  
  can  be used to search for  potential    {\it factorization - breaking} effects in the back-to-back region, 
which  depend 
 on the different color and spin structure of the final states and their interferences with the initial states. 
In order to  investigate these effects experimentally, we propose to measure the ratio of the 
distributions in $\Delta\phi$ for \Zjet - and multijet production at low and at  high transverse momenta, and 
compare the results to predictions obtained assuming factorization.  We examine the role of  
theoretical uncertainties  by performing variations of   the
factorization scale, renormalization scale and matching scale. In particular, we present 
a comparative study of  matching scale uncertainties in the cases of  \PBM -TMD  and 
 collinear parton showers.

\end{abstract}

\section{Introduction} 
\label{Intro}
The description of  jet production in association with a \PZ boson  
in hadron-hadron collisions is an important test of predictions obtained in Quantum Chromodynamics (QCD), and provides 
a relevant background to Higgs boson  studies and to new physics searches. 
The associated  \PZ\ boson plus  jet production has been measured by CDF and D0 in proton-antiproton collisions at a center-of-mass energy $\sqrt{s} = 1.96$~\TeV \cite{Aaltonen:2007ae,Abazov:2008ez}. At the LHC, the  ATLAS and CMS collaborations have published measurements in proton-proton ($\Pp\Pp$)  collisions at a center-of-mass energy $\sqrt{s} = 7$~ \TeV~\cite{Aad:2013ysa,Aad:2011qv,Khachatryan:2014zya}, 8~\TeV~\cite{Khachatryan:2016crw} and 13~\TeV~\cite{Aaboud:2017hbk,Sirunyan:2018cpw}.
Azimuthal correlations  between \PZ~bosons and jets have been measured at 8~\TeV~\cite{Khachatryan:2016crw} and 13~\TeV~\cite{Sirunyan:2018cpw}. 
 
 The distribution in the azimuthal angle $\Delta\phi$ between the \PZ\ boson and the jet  is 
 an especially sensitive observable, probing several aspects of QCD physics. 
At leading order in the strong coupling \alphas , one has  $\Delta\phi= \pi$. 
The smearing of this delta-like distribution is 
a measure of higher order QCD radiation. In the region near $\Delta\phi= \pi$, this is 
primarily soft gluon radiation, while in the region of small  $\Delta\phi$ it is 
primarily hard QCD radiation.  The large-$\Delta\phi$ region of nearly back-to-back 
 \PZ\ boson and  jet  is influenced by both  perturbative and non-perturbative 
QCD contributions. The relative significance of these contributions 
depends on the scale of the transverse momentum imbalance between 
the boson and the jet. Importantly, 
 the resummation of soft multi-gluon emissions in the nearly back-to-back 
 region probes the transverse momenta of the initial state partons, which can 
 be described by 
  transverse momentum dependent (TMD) \cite{Angeles-Martinez:2015sea} parton distribution functions (PDFs). Theoretical predictions for  \PZ\ boson + jet production 
 including soft gluon resummation 
 have recently been given in 
 Refs.~\cite{Bouaziz:2022vp,Chien:2022wiq,Buonocore:2021akg,Chien:2020hzh,Chien:2019gyf,Buffing:2018ggv,Sun:2018icb}.

All the experimental measurements of boson-jet azimuthal correlations 
that have been performed so far are in the kinematical range of transverse 
momenta of the \PZ\ boson and the jets of the order $\pt \approx {\cal O}(100)$ \GeV .  
In this kinematical range, fixed-order perturbative  corrections beyond next-to-leading 
order (NLO) are sizeable, and  at small $\Delta\phi$ 
 NLO calculations are usually not sufficient for reliable predictions.  For 
   the  large-$\Delta\phi$ region of nearly back-to-back  \PZ\ boson and  jet, 
the boson-jet \pt\ imbalance scale is  of order a few GeV,  which is 
 significantly influenced  by both perturbative  resummation and 
 non-perturbative effects. 
It is worth noting that all the experimental measurements performed up to now 
 do not cover 
  the large $\Delta\phi$, nearly back-to-back, region with sufficiently fine binning to investigate detailed features of QCD. 
  
With the increase in luminosity at the LHC, it becomes possible to 
measure \Zjet\ production in the high \pt\ range, 
with $\pt \approx {\cal O}(1000) $  \GeV . In this work,  
we observe that in this kinematical 
range the 
resummation of soft gluons and TMD dynamics in the nearly 
back-to-back region can be explored in a new 
regime,  characterized by   
boson-jet $\pt$ imbalance scales on the order of a few ten GeV.  
The large-$\Delta\phi$  region, involving  deviations of the order of the 
experimental angular resolution of about 1 degree from $\Delta\phi= \pi$,  
can  be investigated 
by analyzing jets with measurable transverse momenta.  

Based on the above observation,  
in this paper we propose experimental investigations of back-to-back 
 azimuthal correlations in the $\pt \approx {\cal O}(1000) $ GeV region, 
with a systematic scan of the large-$\Delta\phi$ regime from this high \pt\  
region down to $\pt \approx {\cal O}(100)$ GeV -- a regime which  is completely 
unexplored experimentally up to now. We present 
dedicated phenomenological studies of this $\Delta\phi$ region  as a function 
of \pt , enabling one to explore boson-jet transverse momentum 
imbalances from a jet scale of several ten GeV down to the few GeV scale. 
To perform these studies, we use the  Parton Branching (\PBM ) 
approach~\cite{Hautmann:2017xtx,Hautmann:2017fcj} to TMD evolution, matched to 
 NLO calculations of   \Zjet\ production   with  \MCatNLO\ \cite{Alwall:2014hca}.
This approach has already been successfully 
applied, across a wide energy and mass range,  
to the  \PZ\ boson \pt\ spectrum  at the LHC \cite{Martinez:2019mwt}  
and the Drell-Yan (DY) \pt\ spectrum  at 
lower fixed-target energies \cite{BermudezMartinez:2020tys}, 
so that the investigation of the same method in the   \Zjet\  case is compelling. 
The $\Delta\phi$ correlation in the kinematical range proposed in this paper 
allows one to study the interplay of 
perturbative and non-perturbative contributions to TMD dynamics 
(see e.g.~\cite{Hautmann:2020cyp} for the DY case) 
as a function of both the boson-jet \pt\ imbalance and 
the evolution scale of the TMD distribution itself, 
of the order of the hard scale of the process, given  
by the transverse momenta of the \PZ\ boson or the jet. 

In a previous publication~\cite{Abdulhamid:2021xtt} we have investigated the \dphi\ correlation in high-\pt\ dijet events by applying 
TMD PDFs and parton shower together with NLO calculations of the hard scattering process. 
In multijet events  the azimuthal correlation between two jets has been measured at the LHC by ATLAS and CMS~\cite{daCosta:2011ni,Khachatryan:2011zj,Khachatryan:2016hkr,Sirunyan:2017jnl,Sirunyan:2019rpc}. 
       The region of $\dphi \to \pi$ is of special interest, since so-called  {\it factorization - breaking}~\cite{Collins:2007nk,Vogelsang:2007jk,Rogers:2010dm} effects could become important in the case of colored final states. 
Multijet production is believed to be sensitive to such effects, 
as well as vector boson + jet production~\cite{Rogers:2013zha}.  
In order to investigate  factorization - breaking effects, we propose to compare the theoretical description of the azimuthal correlation \dphi\ in multijet production with the one in \Zjet\ production. A thorough investigation of azimuthal correlations in the back-to-back region in \Zjet\ events has been also performed in Ref.~\cite{Chien:2022wiq}, addressing the issue of   factorization - breaking.

In this report we compare in detail high-\pt\  dijet  and \Zjet\  production by applying the  \PBM\   TMD method~\cite{Hautmann:2017xtx,Hautmann:2017fcj} matched with NLO.  
In Ref.~\cite{Abdulhamid:2021xtt} 
the NLO  \PBM\ TMD predictions have been found to describe well the 
measurements of  dijet azimuthal correlations~\cite{Sirunyan:2017jnl,Sirunyan:2019rpc}. In the present paper we  apply the same method to 
the calculation of  \Zjet\  production, and present the corresponding predictions. 
We propose to use the same kinematic region for the  high-\pt\  dijet  and \Zjet\ production to allow a direct comparison of the angular observables in the two cases. 

We will see that, in the region of leading transverse momenta 
of the order $\pt \approx {\cal O}$(100 GeV), the boson-jet final state is more 
strongly correlated azimuthally than the jet-jet final state. As the  
transverse momenta increase above the electroweak symmetry breaking scale, 
 $\pt \approx {\cal O}$(1000 GeV),  this difference is 
 reduced, and the boson-jet and jet-jet become more 
 similarly correlated. We connect  this behavior  to features of the partonic 
 initial state and final state radiation  in the boson-jet and jet-jet cases. 
Since  potential factorization-breaking effects arise from 
 color interferences of initial-state and final-state radiation, 
 different  breaking patterns can be expected for  
 strong and weak azimuthal correlations, 
 influencing differently the boson-jet and jet-jet cases.  We therefore propose 
 to systematically compare measurements of 
 dijet  and \Zjet\ distributions, scanning the 
 phase space  from low transverse momenta $\pt \approx {\cal O}$(100 GeV) 
 to high transverse momenta   $\pt \approx {\cal O}$(1000 GeV). 

In the following,  we start by describing the basic elements 
of the  \PBM\   TMD method   and 
the  \Zjet\  calculation in Sec.~2. 
In Sec.~3 we present  results for the   \Zjet\ azimuthal correlations and compare them with the multijet case. 
We summarize in Sec.~4. In an appendix we discuss technical details on the use of \mcatnlo+\cascade .

\section{Basic elements of the calculation}

In this section we first recall the salient features of the  \PBM\   TMD 
approach, summarizing the main concepts of the approach and its  
applications; then we  
  describe the calculation of   \Zjet\  production by the 
 \PBM\   TMD method matched with NLO matrix elements in 
  \MCatNLO . 
 
\subsection{PB - TMD method}

The  \PBM\ approach~\cite{Hautmann:2017fcj}
provides a formulation for the evolution of TMD parton 
distributions 
in terms of perturbatively calculable Sudakov form factors and
real-emission splitting kernels, with angular ordering phase space 
constraints and with non-perturbative distributions at the initial scale of 
the evolution to be determined from fits to experiment. 
This formulation uses a 
soft-gluon resolution scale $z_M$~\cite{Hautmann:2017xtx}  
to separate resolvable and non-resolvable branchings. 
An important feature of the  \PBM\   TMD evolution equation~\cite{Hautmann:2017fcj}   
concerns its collinear limits: upon integration over all transverse 
momenta,  the  \PBM\   TMD evolution equation 
returns the DGLAP~\cite{\dglap}  equation  for  resolution scale $z_M \to 1$, 
while it coincides with the CMW~\cite{Marchesini:1987cf,Catani:1990rr} coherent branching equation for angular-ordered $z_M$~\cite{Hautmann:2019biw}. 
The  \PBM\   TMD method 
is based on the ``unitarity'' picture~\cite{Webber:1986mc} 
of parton evolution usually employed in parton showering Monte Carlo (MC) 
algorithms~\cite{Bellm:2015jjp,Sjostrand:2014zea}. The  \PBM\ 
evolution equation 
for the TMD distributions is matched by a corresponding TMD parton shower 
for the spacelike parton cascade,   
generated by ``backward evolution''~\cite{Baranov:2021uol}.  
A significant difference with respect to ordinary parton showers 
is that in the  \PBM\   TMD method  
TMD distributions are defined and determined from fits to experimental 
   data, which places constraints on fixed-scale inputs
to evolution, while in ordinary parton showers 
instead nonperturbative
physics parameters and showering parameters are tuned. No MC tuning is performed in the 
  \PBM\   TMD case.  

The NLO \PBM\  collinear and TMD parton distributions were 
obtained in Ref.~\cite{Martinez:2018jxt} from QCD fits to precision DIS data from 
HERA~\cite{Abramowicz:2015mha} using the xFitter analysis framework \cite{xFitterDevelopersTeam:2022koz,Alekhin:2014irh}. Two different sets, PB-NLO-2018-Set~1 and PB-NLO-2018-Set~2, were obtained, with PB-NLO-2018-Set~1 corresponding at collinear level to HERAPDF 2.0 NLO~\cite{Abramowicz:2015mha}. In PB-NLO-2018-Set~2 the transverse momentum (instead of the evolution scale in Set~1) is used as the scale in the running coupling $\alphas$ which corresponds to the angular ordering of soft gluon emissions in the initial-state parton evolution~\cite{Bassetto:1983mvz,Dokshitzer:1987nm,Catani:1990rr,Hautmann:2019biw}. It has been shown 
in~\cite{BermudezMartinez:2020tys,Martinez:2019mwt}  that Set~2 provides a better description of experimental measurements for the  \PZ\ - boson  spectrum at  
low-\pt .  Also, it has been shown in~\cite{Abdulhamid:2021xtt} that 
the transverse momentum scale in the running coupling  \alphas\  is important 
for a good / of data on di-jet angular correlations. 
  In this paper we will concentrate on Set 2 only.

\begin{figure}[h!tb]
\begin{center} 
\includegraphics[width=0.45\textwidth]{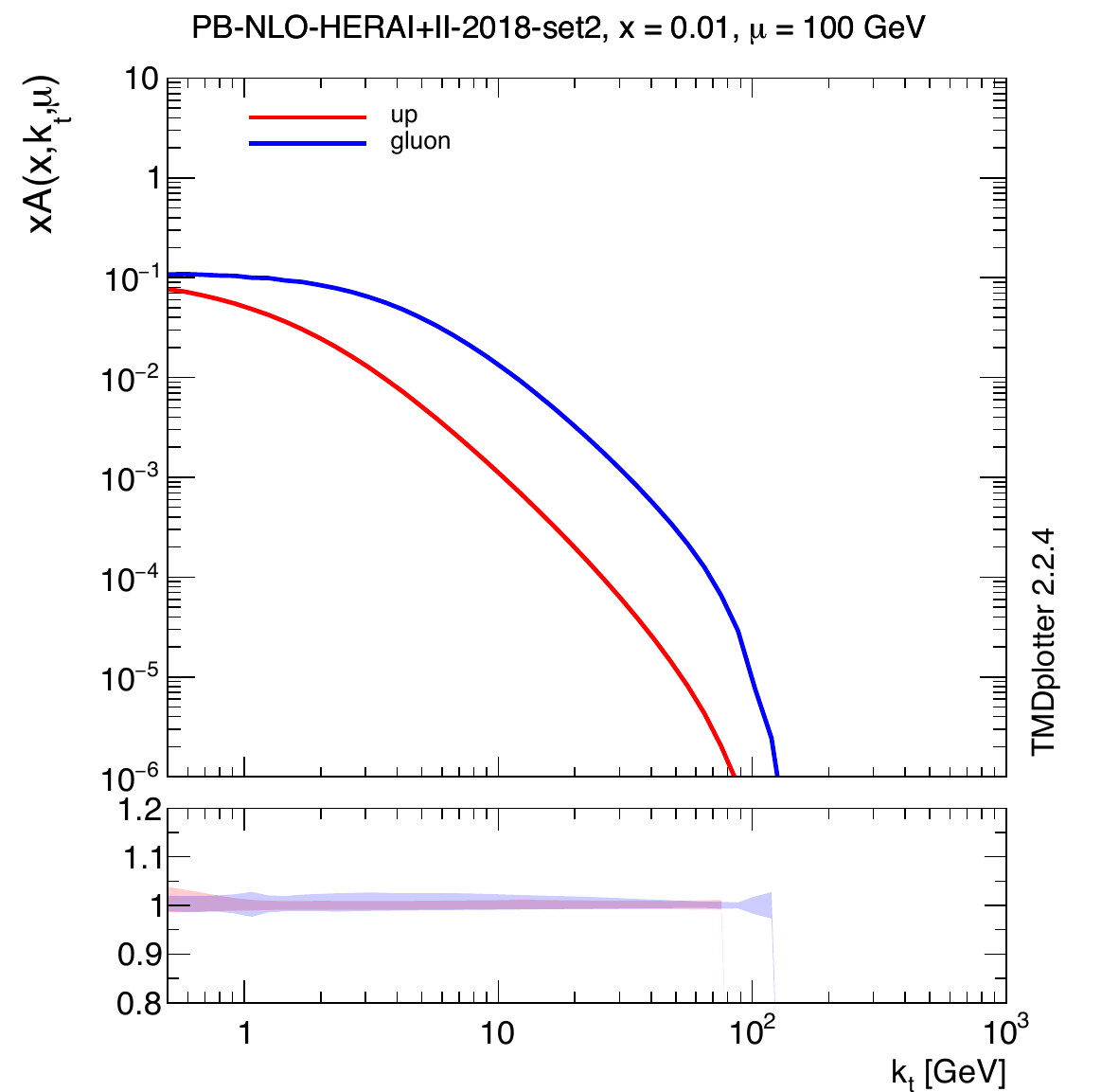} 
\includegraphics[width=0.45\textwidth]{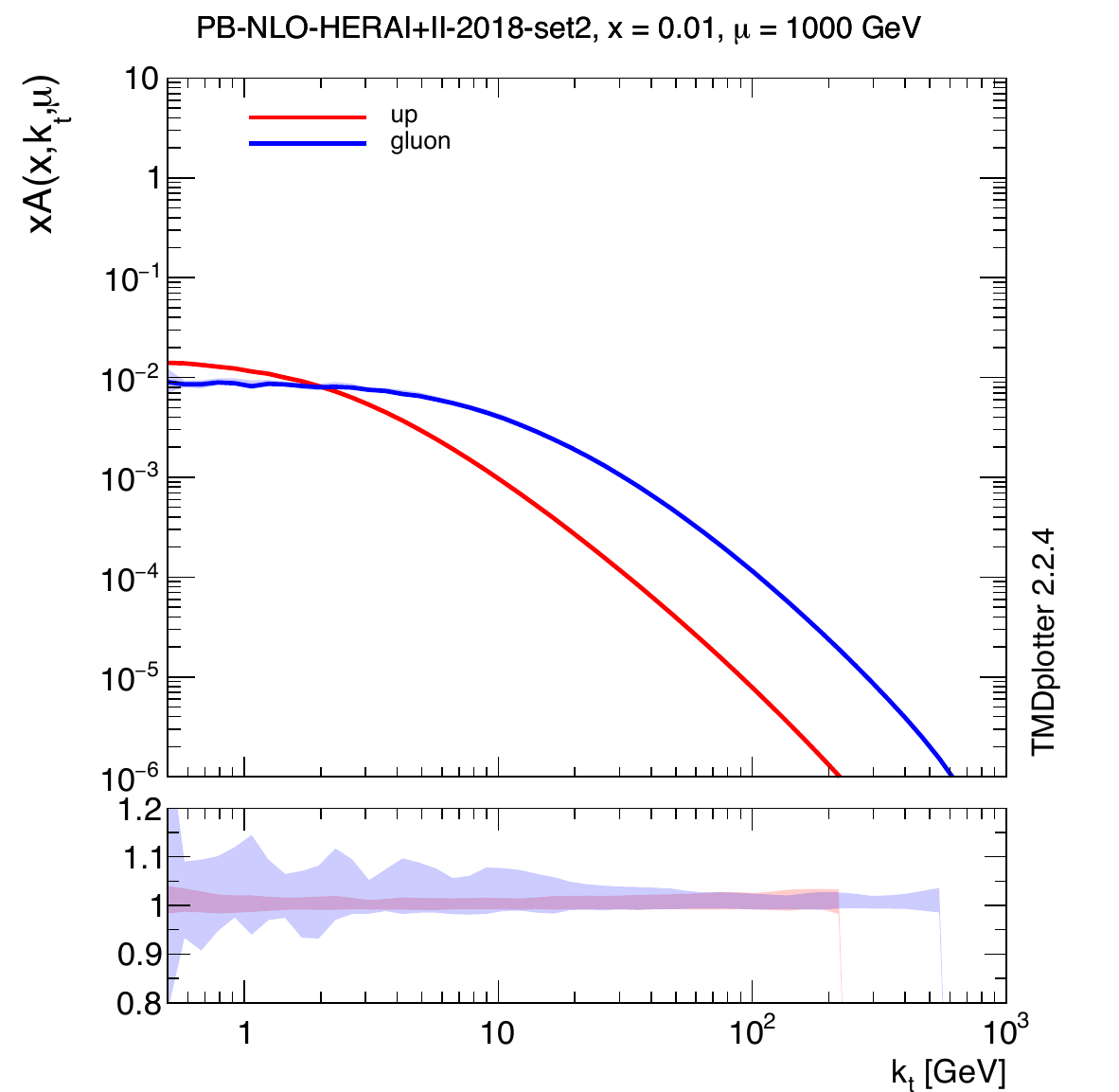} 
  \caption{\small TMD parton density distributions for up quarks and gluons of PB-NLO-2018-Set~2 as a function of $\kt$ at $\mu=100$ and $1000$  \GeV\  and $x=0.01$.
  In the lower panels show the full uncertainty of the TMD PDFs, as obtained from the fits \protect\cite{Martinez:2018jxt}.
  }
\label{TMD_pdfs}
\end{center}
\end{figure}

In Fig.~\ref{TMD_pdfs} we show the TMD PDF distributions for up quarks and gluons at $x=0.01$ and $\mu=100$ and $1000$ \GeV\ for PB-NLO-2018-Set~2. The transverse momentum distribution of gluons is broader than that of quarks, due to gluon self-coupling and the different color factors. In Fig.~\ref{TMD_pdfs} also the uncertainties of the distributions, as obtained from the fit~\cite{Martinez:2018jxt}, are shown. The differences in the transverse momentum spectra of quarks and gluons will show up in differences in azimuthal correlation distributions.

The  \PBM\   TMD evolution equation resums Sudakov logarithms. Current 
calculations in the  \PBM\   TMD approach 
are performed with leading-logarithm (LL) 
and next-to-leading-logarithm  (NLL) accuracy. The accuracy can be 
systematically improved, and the  extension to 
next-to-next-to-leading logarithmic (NNLL) accuracy is being studied. 
In this respect, the approach can be compared~\cite{vanKampen:2021oxe} 
 with analytic 
resummation methods~\cite{Collins:1984kg,Collins:2011zzd}.  
The extraction of TMD distributions 
from the  \PBM\   TMD fits described above could be compared 
with extractions, such as~\cite{Bacchetta:2022awv,Bury:2022czx}, based 
on~\cite{Collins:1984kg,Collins:2011zzd}. The TMDlib  
tool~\cite{Abdulov:2021ivr,Hautmann:2014kza} is 
designed as an aid for such studies.  On the other hand, while 
analytic resummation approaches apply to the inclusive 
transverse momentum spectrum, the  \PBM\   TMD approach 
works at exclusive level and can be 
applied to make predictions not only for the inclusive spectrum 
but also for the structure of the final states. 

A framework to compute theoretical predictions 
combining the   \PBM\   TMD resummation with fixed-order NLO matrix 
elements in   \MCatNLO\ has been developed 
in~\cite{Martinez:2019mwt,BermudezMartinez:2020tys}. 
The predictions~\cite{Martinez:2019mwt} 
have been successfully compared with 
LHC measurements of  \PZ~boson \pt\ and 
$\phi^*$ distributions~\cite{Aad:2015auj,Sirunyan:2019bzr,CMS:2022ubq}. 
Predictions by this method have also been successful in 
describing~\cite{BermudezMartinez:2020tys} DY  \pt\ 
spectra at lower masses and 
energies~\cite{Aidala:2018ajl,Antreasyan:1981eg,Webb:2003ps,Webb:2003bj}.   
The significance of this result  is enhanced by the recent 
observation~\cite{Gauld:2021pkr} that fixed-order NNLO corrections 
are not extremely large in the kinematic region of the data. 
This framework has also been applied  to 
di-jet production~\cite{Abdulhamid:2021xtt}, and 
predictions for di-jet correlations have been found 
in good agreement with LHC 
measurements~\cite{Sirunyan:2017jnl,Sirunyan:2019rpc}.  
We will employ this framework for 
  \Zjet\  production in the next subsection. 

As a method which is applicable at the level of 
exclusive final states, 
the  \PBM\   TMD approach can be used in the context 
of multi-jet merging algorithms. A TMD multi-jet merging 
method has been 
developed in~\cite{Martinez:2021chk}.  
Its application  to 
 \PZ~boson + multi-jets production~\cite{Martinez:2022wrf,Martinez:2021chk,Martinez:2021dwx} 
illustrates that transverse momentum recoils in 
the initial-state 
showers~\cite{Hautmann:2013fla,Dooling:2012uw,Hautmann:2012dw}  
 influence significantly the theoretical systematics associated with the 
merging parameters.   
In the present paper, we will concentrate on the   \Zjet\   back-to-back region, 
rather than the multi-jet production region, and we will therefore not use the 
TMD merging procedure. 

Recently, the  \PBM\   TMD  evolution equation has been generalized 
to include TMD splitting 
functions~\cite{Hautmann:2022xuc,Keersmaekers:2021arn}, 
defined through high-energy factorization~\cite{Catani:1994sq}. 
This generalization is important particularly for processes sensitive to 
TMD distributions at small values of longitudinal momentum fractions $x$.
In this paper we focus on  processes at mid to large $x$,  
and thus we do not consider this in the following.

\subsection{Calculation of Z+jet  distributions}

 \begin{tolerant}{1000} 
The process \Zjet\ at NLO is calculated with \MCatNLO\ using the collinear PB-NLO-2018-Set 2, as obtained in Ref.~\cite{Martinez:2018jxt} applying  $\alphas(M_{\PZ } )= 0.118$.  The matching of NLO matrix elements with  \PBM\ TMD parton distributions is described in Refs.~\cite{Martinez:2019mwt,BermudezMartinez:2020tys,Baranov:2021uol}. The extension to multijet production is illustrated in Ref.~\cite{Abdulhamid:2021xtt}. Predictions are obtained by processing the \MCatNLO\ event files in LHE format~\cite{Alwall:2006yp} through \cascade~\cite{Baranov:2021uol} for an inclusion of TMD effects in the initial state and for  simulation of the corresponding parton shower (labeled \mcatnlo+CAS3\ in the following).  
\end{tolerant}

\begin{tolerant}{1000}
Fixed order NLO \Zjet\ production is calculated with \MCatNLO\, in a procedure similar to the one applied for dijet production described in~\cite{Abdulhamid:2021xtt} (labeled \mcatnlo (fNLO)). For the \mcatnlo\ mode, the  \herwig6  \cite{Corcella:2002jc,Marchesini:1991ch}  subtraction terms are calculated, as they are best suited for the use with \PBM\ - parton densities, because both apply the same angular ordering condition. The use of \herwig6 subtraction terms together with \cascade\ is justified in appendix Section~\ref{appendix1} for final state parton shower as well as initial and final state showers by a comparison of the predictions obtained with \cascade\ and with \herwig6.
The matching scale $\mu_{m}=\verb+SCALUP+$ limits the contribution from \PBM -TMDs and TMD showers.
\end{tolerant}

\begin{tolerant}{1000}
In the calculations, the factorization and renormalization scales are set to  $\mu=\frac{1}{2} \sum_i p_{{\rm T},i} $, where the index $i$ runs over all particles in the matrix element final state. This scale is also used in the \PBM -TMD parton distribution ${\cal A}(x,\kt,\mu)$. The scale uncertainties of the predictions are obtained from variations of the scales around the central value in the 7-point scheme avoiding extreme cases of variation.
\end{tolerant}

\begin{figure}[h!tb]
\begin{center} 
\includegraphics[width=0.48\textwidth]{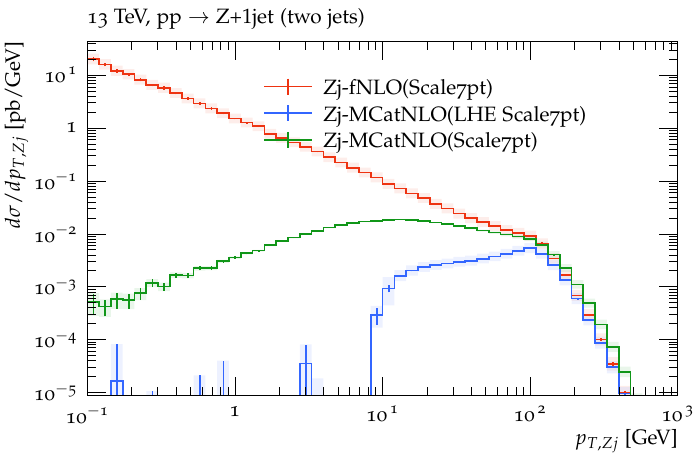} 
\includegraphics[width=0.48\textwidth]{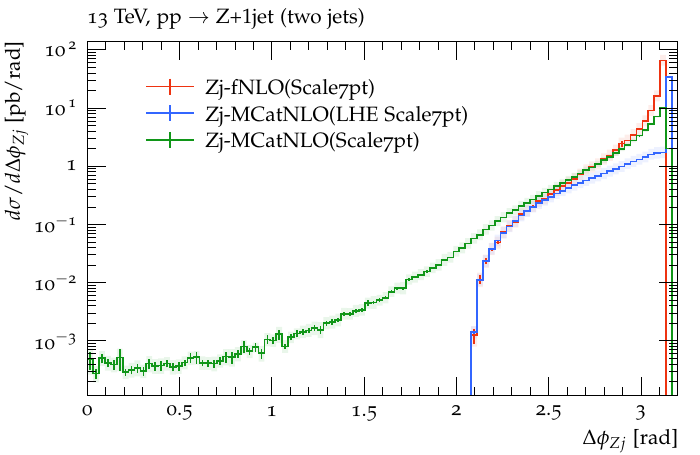} 
  \caption{\small Transverse momentum spectrum of the \Zjet -system \ptZj\ (left)  and $\dphiZ$ distribution (right).
Shown are predictions from fixed NLO (fNLO), the (unphysical) distribution at LHE-level and the full simulation (after inclusion of \PBM -TMDs and TMD showers, \mcatnlo+CAS3). }
\label{mcatnlo_lhe}
\end{center}
\end{figure} 

In Fig.~\ref{mcatnlo_lhe} we show the distributions of the transverse momentum of the \Zjet\ system, \ptZj ,  and the azimuthal correlation in the \Zjet\ system, \dphiZ , for a fixed NLO calculation, for the full simulation including \PBM -TMD PDFs and parton showers as well as for the \mcatnlo\ calculation at the level where subtraction terms are included without addition from parton shower (LHE-level). We require a transverse momentum $\pt > 200 $ \GeV\ for the \PZ boson and define
jets with the anti-\kt\ jet-algorithm~\cite{Cacciari:2008gp}, as implemented in the FASTJET package~\cite{Cacciari:2011ma}, with a distance parameter of R=0.4. The effect of including \PBM -TMD PDFs and parton showers can be clearly seen from the difference to the fixed NLO and LHE-level calculations.

In the low \ptZj\ region one can clearly see the expected steeply rising behavior of the fixed NLO prediction. In the \dphiZ\ distribution one can observe the limited region for fixed NLO at $ \dphiZ  < 2/3 \pi$, since at most two jets in addition to the \PZ boson appear in the calculation. At large \dphiZ ,  the fixed NLO prediction rises faster than the full calculation including resummation via \PBM -TMDs and parton showers. In the following we concentrate on the large \dphiZ\ region.

\section{Back-to-back azimuthal correlations in  \Zjet\  and multijet production}
\label{sec:correlations}
We now present predictions, obtained in the framework described above, for \Zjet\ and multijet 
production.\footnote{A   framework based on CCFM evolution~\cite{Jung:2010si} 
was described in~\cite{Dooling:2014kia,Hautmann:2008vd} for multi-jet and vector boson + jet correlations.} 
The selection of events follows the one of  azimuthal correlations \dphi\ in the back-to-back region ($\dphi \to \pi$) in multijet production at $\sqrt{s}=13$~\TeV\  as obtained by CMS~\cite{Sirunyan:2019rpc}: 
jets are reconstructed with the anti-\kt\ algorithm~\cite{Cacciari:2008gp} with a distance parameter of~0.4 in the rapidity range of $|y|< 2.4$.
We require either two jets with $\ptmax > 200$~\GeV\ or 
a \PZ\ boson and a jet as leading or subleading objects with a transverse momentum $\ptmax>200$~\GeV . 

We consider distributions of the azimuthal correlation between the \PZ boson and the leading jet, \dphiZ , for $\ptmax > 200$ \GeV\ as well as for the very high \pt\ region of $\ptmax > 1000$ \GeV . 

The calculations are performed with \mcatnlo+CAS3 using  PB-NLO-2018-Set~2 as the collinear and TMD parton densities with running coupling satisfying $\asmz = 0.118$ and \PBM -TMD parton shower.

\begin{figure}[h!tb]
\begin{center} 
\includegraphics[width=0.45\textwidth]{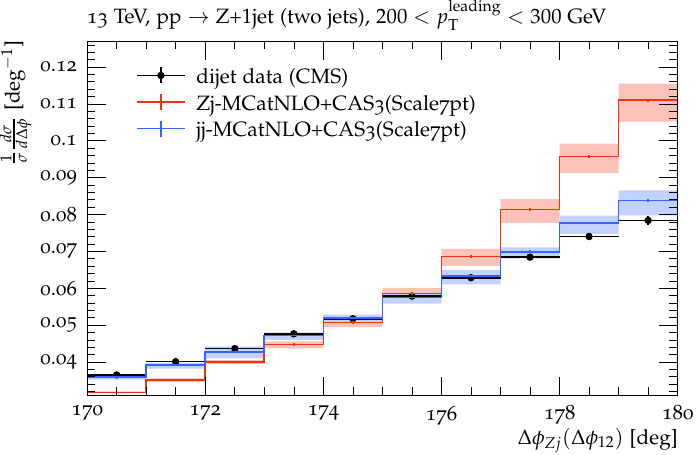} 
\includegraphics[width=0.45\textwidth]{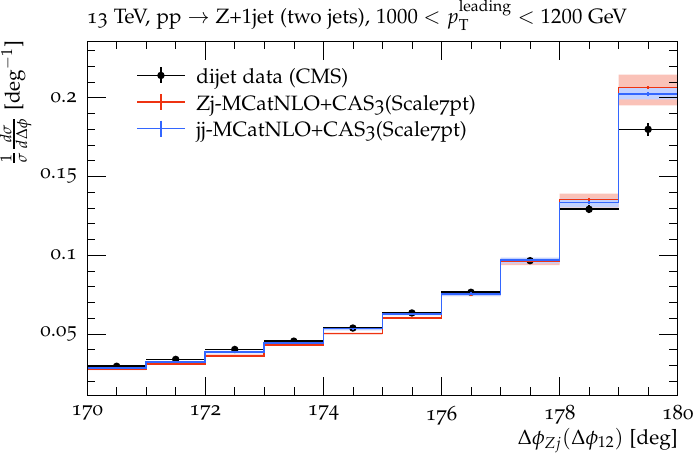} 
  \caption{\small Predictions of the azimuthal correlation \dphiZ (\dphi) for \protect\Zjet\ and multijet processes in the back-to-back region for $\ptmax > 200 $ \GeV\ (left) and $\ptmax > 1000 $ \GeV\ (right)  obtained from \protect\mcatnlo+CAS3. 
  Shown are the uncertainties obtained from scale variation (as described in the text).
  The measurements of dijet correlations as obtained by CMS~\protect\cite{Sirunyan:2019rpc} are shown as 
   data points, for comparison. 
}
\label{b2b-Zets_CAS}
\end{center}
\end{figure}

In Fig.~\ref{b2b-Zets_CAS}, the prediction for the azimuthal correlations \dphiZ\  for \Zjet\ production in the back-to-back region is shown.\footnote{Predictions for the region of small $\Delta \phi$  require including the contribution of 
higher parton multiplicities, e.g.~via  multi-jet merging~\cite{Martinez:2021chk}.}   
We also show, for comparison, the prediction of azimuthal correlations \dphi\ for multijet production in the same kinematic region, compared to the measurement of dijet production obtained by CMS~\cite{Sirunyan:2019rpc}. 
We observe that the distribution of azimuthal angle \dphiZ\ in \Zjet -production for $\ptmax >200$ \GeV\  is more strongly correlated towards $\pi$ than the distribution of angle \dphi\ in multijet production. This difference is reduced for $\ptmax > 1000$ \GeV . 

Differences in $\Delta \phi$  between \Zjet\ and multijet production can result from the different flavor composition of the initial state and therefore different initial state transverse momenta  and initial state parton shower, as well as from differences in final state showering since both processes have a different number of colored final state  partons. Effects coming from { factorization - breaking}, interference between initial and final state partons,  will  depend on the final state structure and the number of colored final state partons.

\begin{figure}[h!tb]
\begin{center} 
\includegraphics[width=0.32\textwidth]{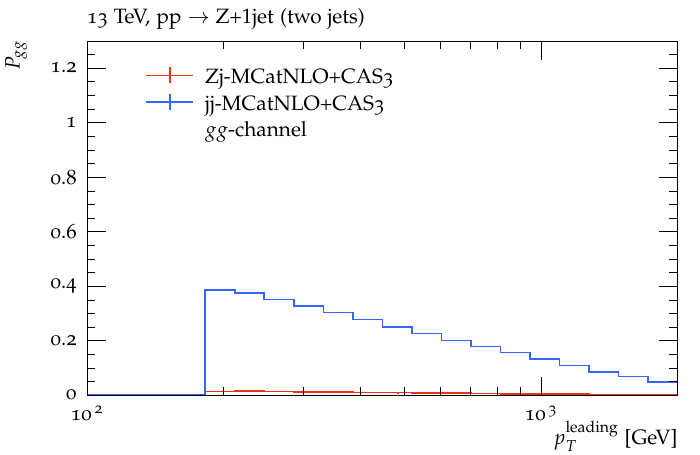} 
\includegraphics[width=0.32\textwidth]{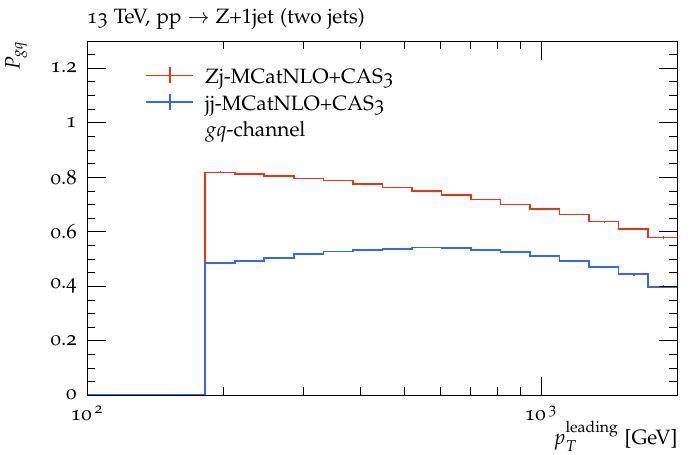} 
\includegraphics[width=0.32\textwidth]{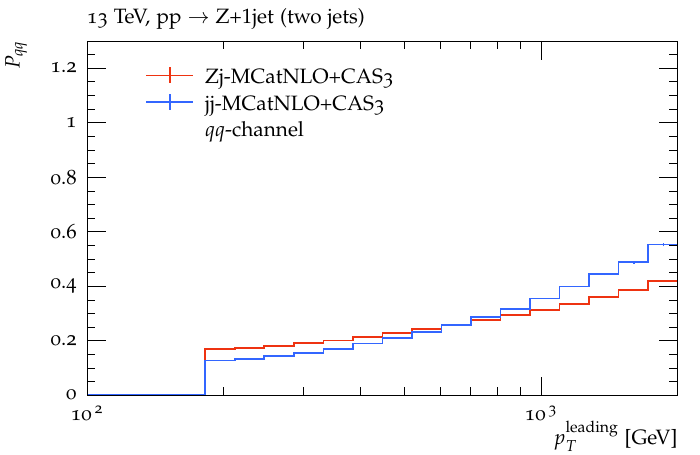} 
  \caption{\small The probability of $gg$, $qg$ and $qq$ initial states in \Zjet\ and multijet production ($q$ stands for quark and antiquark) as a function of \protect\ptmax . 
The predictions are  calculated with  \protect\mcatnlo+CAS3. }
\label{flavor_compostion}
\end{center}
\end{figure}

We first investigate the role of initial state radiation and the dependence on the transverse momentum distributions coming from the TMD PDFs,
which gives a large contribution to  the decorrelation in $\Delta \phi$. The \kt -distribution obtained from a gluon TMD PDF is different from the one of a quark TMD PDF as shown in Fig.~\ref{TMD_pdfs} for $x=0.01$ and scales of $\mu=200 (1000) $ \GeV .
In Fig.~\ref{flavor_compostion}  we show the probability of $gg$, $qg$ and $qq$ initial states ($q$ stands for quark and antiquark) as a function of \ptmax\ for \Zjet\ and multijet production obtained with \mcatnlo+CAS3.  At high $\ptmax > 1000$ \GeV\ the $qq$ channel becomes important for both \Zjet\ and multijet final states, while at lower $\ptmax > 200$ \GeV\ the $gg$ channel is dominant in multijet production, leading to larger decorrelation effects, since gluons radiate more compared to quarks.

\begin{figure}[h!tb]
\begin{center} 
\includegraphics[width=0.45\textwidth]{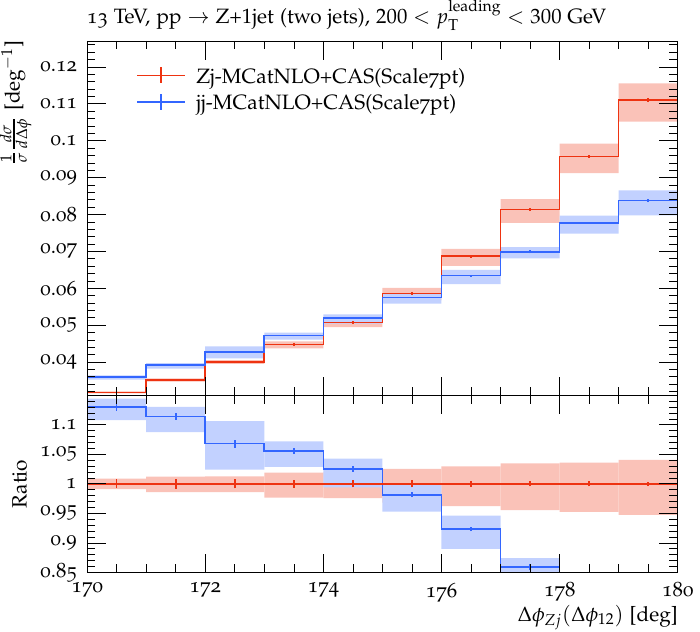} 
\includegraphics[width=0.45\textwidth]{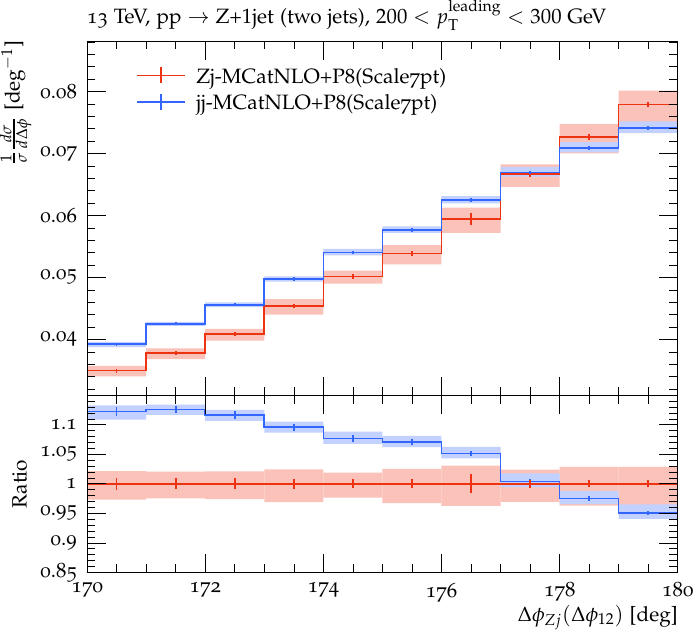} 
\includegraphics[width=0.45\textwidth]{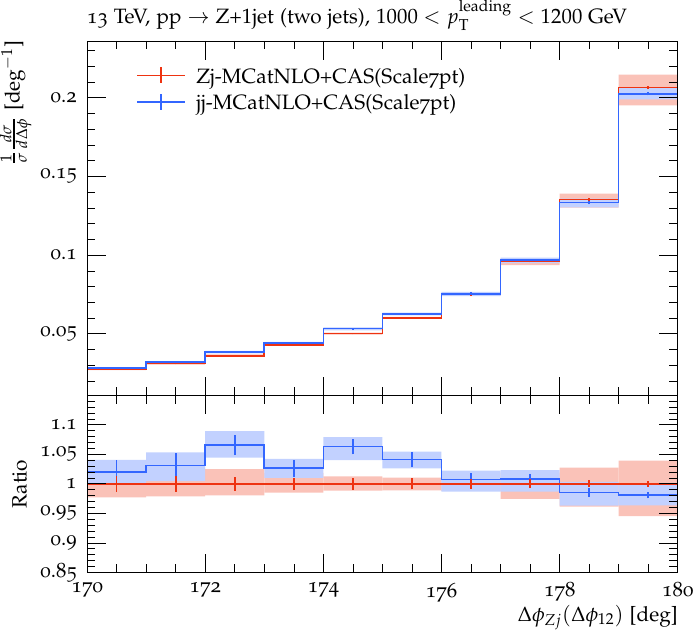} 
\includegraphics[width=0.45\textwidth]{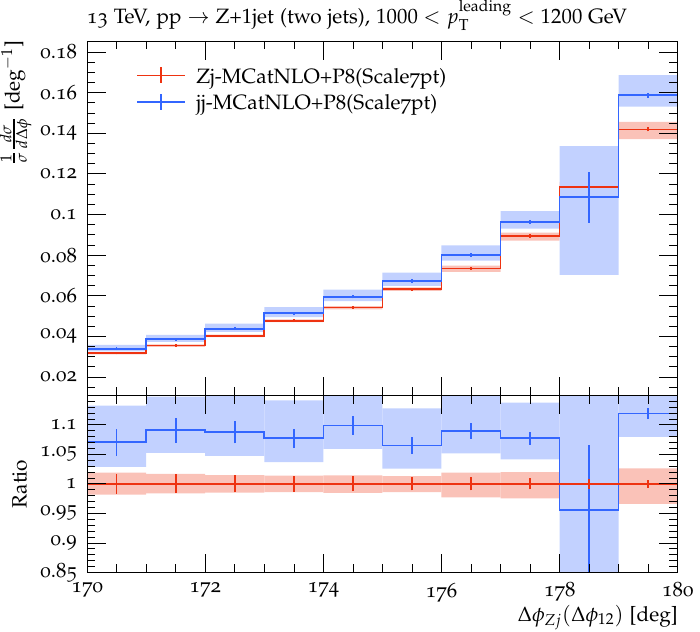} 
  \caption{\small Predictions for the azimuthal correlation \dphiZ (\dphi)  in the back-to-back region for \protect\Zjet\ and multijet production obtained with \protect\mcatnlo+CAS3 (left column) and  \protect\mcatnlo+\pythia 8 (right column). Shown are different regions in  $\ptmax > 200 $~\GeV\ (upper row) and $\ptmax > 1000 $~\GeV\ (lower row). The bands show the uncertainties obtained from scale variation (as described in the text).  }
\label{b2b-Zets_CAS_P8}
\end{center}
\end{figure}

\begin{tolerant}{1000}
The role of final state radiation in the correlation in \dphi\ distributions is more difficult to estimate, since the subtraction terms for the NLO matrix element calculation also depend on the structure of the final state parton shower. In order to estimate the effect of final state shower we compare a calculation of the azimuthal correlations in the back-to-back region obtained with \mcatnlo+CAS3 with the one obtained with \mcatnlo+\pythia 8 (Fig.~\ref{b2b-Zets_CAS_P8}). For the calculation \mcatnlo+\pythia 8 we apply the  \pythia 8 subtraction terms in the \MCatNLO\ calculation, use the NNPDF3.0~\cite{Ball:2014uwa} parton density and tune CUETP8M1~\cite{Khachatryan:2015pea}.
\end{tolerant}

As shown in Fig.~\ref{b2b-Zets_CAS_P8}, the distributions are different because of the different  parton shower in \cascade\ and \pythia 8, but the ratio of the distributions for \Zjet\ and multijet production are similar: \Zjet -production gives a steeper (more strongly correlated) distribution at low \ptmax , while at high \ptmax\ the distributions become similar in shape.  We conclude, that the main effect of the $\Delta\phi$ decorrelation comes from initial state radiation, and the shape of the $\Delta\phi$ decorrelation in the back-to-back region becomes similar between \Zjet\ and dijet processes at high \ptmax\ where similar initial  partonic states are important.

\begin{figure}[h!tb]
\begin{center} 
\includegraphics[width=0.45\textwidth]{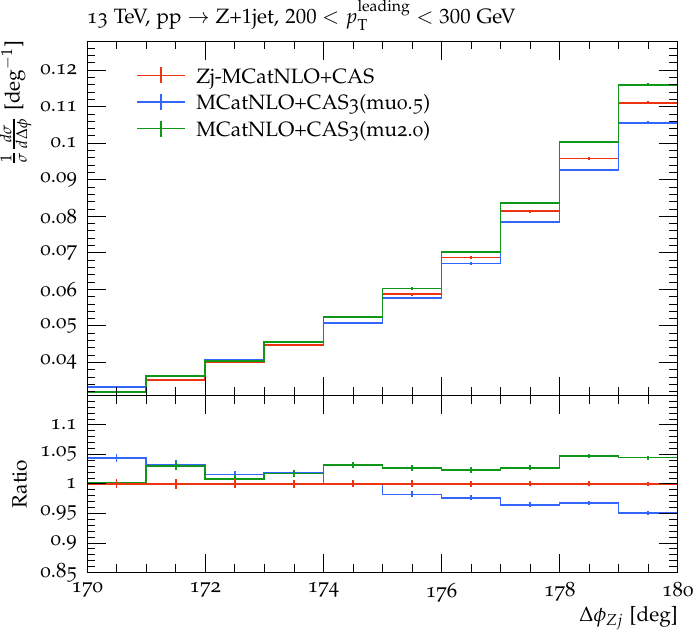} 
\includegraphics[width=0.45\textwidth]{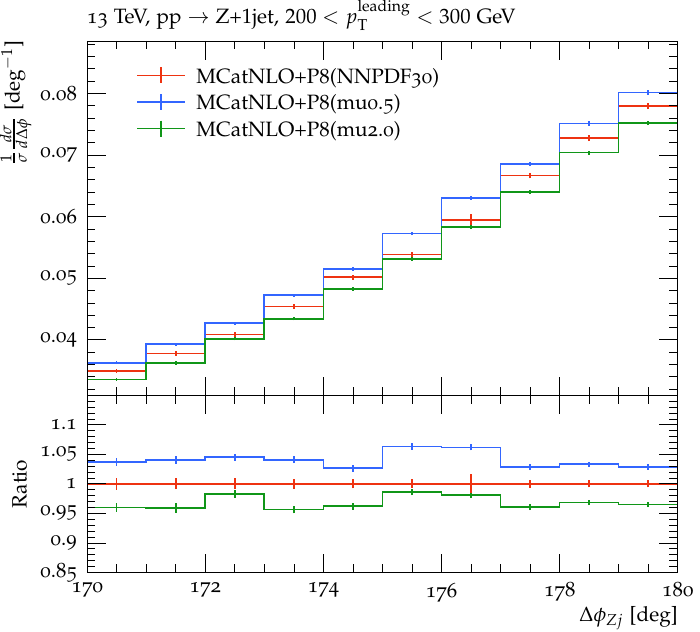} 
\includegraphics[width=0.45\textwidth]{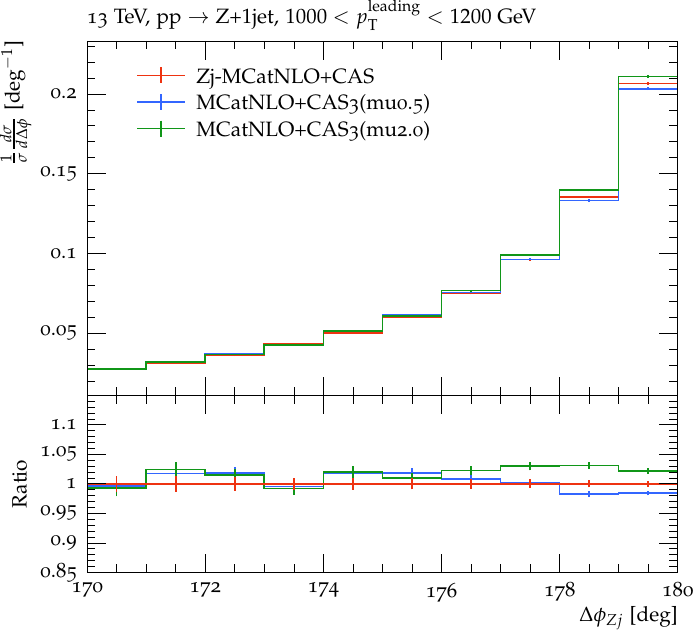} 
\includegraphics[width=0.45\textwidth]{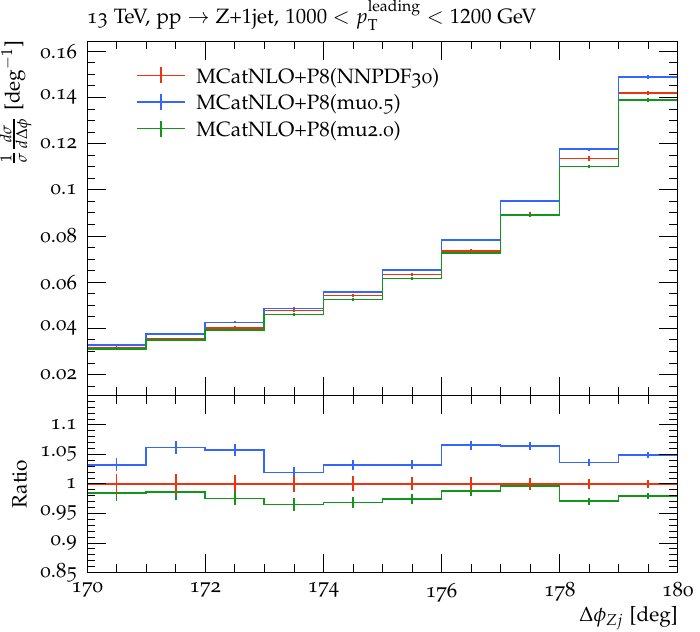} 
  \caption{\small The dependence on the variation of the matching scale $\mu_{m}$ in
  predictions for the azimuthal correlation \dphiZ (\dphi)  in the back-to-back region.
  Shown are predictions  obtained with \protect\mcatnlo+CAS3 (left column) and  \protect\mcatnlo+\pythia 8 (right column)
  for $\ptmax > 200 $ \GeV\ (upper row) and $\ptmax > 1000 $ \GeV\ (lower row).  
  The predictions with different matching scales $\mu_{m}$ varied by a factor of two up and down are shown.}
\label{b2b-Zets_CAS_P8_mum}
\end{center}
\end{figure}

\begin{tolerant}{1000}
The matching scale $\mu_{m}$ limits the hardness of parton-shower emissions, and is thus typically a non-negligible source of variation in matched calculations (see e.g.\ \cite{Bellm:2016rhh} for a detailed discussion). It is thus interesting to assess the robustness of the previous findings under variations of the matching scale. Assessing matching scale variations in both an angular-ordered shower -- such as \cascade\ -- and a transverse-momentum-ordered shower  -- such as \pythia 8 -- additionally tests the \emph{interpretation (role)} of the matching scale. In transverse-momentum ordered showers, the matching scale sets the maximal transverse momentum of the first shower branchings, while branchings beyond the first emission are not explicitly affected by the matching scale. In an angular-ordered shower, however, the matching scale is applied as "veto scale" to avoid larger transverse momenta for any branching, i.e.\ the matching scale directly affects all branchings. The result of changing the matching scale to half or twice the central value is shown in Fig.~\ref{b2b-Zets_CAS_P8_mum}. As expected, the value of the matching scale has an impact on the prediction  ($\sim 5\%$). This is particularly apparent when $\mu_m$ is used to set the maximal transverse momentum of the first emission in \pythia 8. Overall, we find that interpreting the matching scale as veto scale in \cascade\ leads to apparently more robust predictions. Interestingly, the matching scale uncertainty becomes smaller for higher-\ptmax\ jet configurations in \cascade . The size of the matching scale variation is comparable to scale variations, and should thus be carefully studied when designing uncertainty estimates. 
\end{tolerant}

In dijet production the measurements are rather well described with predictions obtained with \mcatnlo+CAS3, as shown in Fig.~\ref{b2b-Zets_CAS}  and discussed in detail in Ref.~\cite{Abdulhamid:2021xtt}. Only in the very high \ptmax\ region, a deviation from the measurement is observed, which could be perhaps interpreted as coming from a violation of factorization. It is therefore very important to measure $\Delta\phi$ distributions in other processes, where factorization is expected to hold.

In order to experimentally probe effects which could originate from {factorization - breaking} in the back-to-back region we propose to measure the ratio of distributions in \dphiZ\ for \Zjet\  and  \dphi\ for multijet production at low and very high \ptmax , and compare the measurement with predictions assuming that factorization holds. The number of colored partons involved in \Zjet\ and multijet events is different, and deviations from factorization will depend on the structure of the colored initial and final state. In order to minimize the effect of different initial state configurations, a measurement at high \ptmax ,  hint more clearly at possible  {factorization - breaking} effects.

In Ref.~\cite{Chien:2022wiq} a detailed study on \Zjet\ azimuthal correlations is reported, applying TMD-factorization and the "winner-takes-all" jet recombination scheme, with the aim to reduce potential factorization breaking contributions. We have checked  that our main results remain unchanged when the "winner-takes-all" jet recombination scheme \cite{Fastjet-wta,Bertolini:2013iqa} is applied, only in the last bin of the $\dphiZ$ distributions the cross section is reduced. We find that multijet events are more affected by the "winner-takes-all" jet recombination scheme in the back-to-back region at high \pt\ than  \Zjet\ events.

\section{Summary and conclusions}
We have investigated azimuthal correlations in  \Zjet\ production and compared predictions with those for multijet production in the same kinematic range. The predictions are based on  \PBM -TMD distributions with  NLO calculations via \mcatnlo\ supplemented by \PBM -TMD parton showers via \cascade . The azimuthal correlations \dphiZ , obtained in \Zjet\  production are steeper compared to those in multijet production (\dphi ) at transverse momenta ${\cal O}(100)$ \GeV , while they become similar for very high transverse momenta, ${\cal O}(1000)$ \GeV , which is a result of  similar initial parton configuration of both processes.

In \Zjet\ production the color and spin structure of the partonic final state is different compared to the one in multijet production, and differences in the azimuthal correlation patterns can be used to search for  potential 
{factorization - breaking} effects, involving initial and final state interferences. In order to experimentally investigate those effects, we propose to measure the ratio of the distributions in \dphiZ \ for \Zjet - and \dphi\ for multijet production at low and at very high \ptmax , and compare the measurements to predictions obtained assuming that factorization holds. 

We have studied the matching scale dependence in the  \PBM -TMD predictions and compared it with the 
case of NLO-matched calculations based on the  \pythia 8  collinear shower.   We find that variations of the 
matching scale  lead to more stable predictions in the  \PBM -TMD case, with  the 
relative reduction of the matching scale 
theoretical uncertainty becoming more pronounced for increasing \ptmax\ transverse momenta.

\vskip 1 cm 
\begin{tolerant}{8000}
\noindent 
{\bf Acknowledgments.} 

\noindent 
We are grateful to Olivier Mattelaer from the \MCatNLO\ team for discussions, help and support with the lhe option for fixed NLO calculations in \mcatnlo .
\end{tolerant} 
\vskip 0.6cm 

\clearpage

\section{Appendix: Comparison of {\cascade\ } and \herwig6 \label{appendix1}}

The calculations presented here apply the \mcatnlo\ method using \herwig6 (H6) subtraction terms, as implemented in \MCatNLO . The NLO accuracy of the calculations is preserved by construction, since the use of \PBM -TMD distributions and TMD shower, as well as the ordinary parton shower, does not change the inclusive cross section.

Since \herwig6 (H6) subtraction terms are used in the \mcatnlo+CAS3 calculations, we investigate here in detail the contribution of the parton shower used in \cascade . We compare predictions obtained with \mcatnlo+CAS3 with the corresponding ones obtained with \mcatnlo+H6, using  LHE files produced with \MCatNLO\ for \PZ production.
The \PZ boson is reconstructed from two oppositely charged leptons with $\pt > 20 $~\GeV\ in $|\eta|<2.4$. 
We also study jet distributions obtained with the anti-\kt\ algorithm with distance parameter 0.4  with $\pt > 30$ \GeV\ and $|\eta| < 5 $.  

In H6 the allowed region of $z$ for a branching $q \to qg$ in the final state shower is $Q_q/Q < z < 1-Q_g/Q $ (e.g. A.2.2 in Ref.~\cite{Frixione:2003ei}), with $Q_q = m_q + {\tt VQCUT}$ and $Q_g = m_g + {\tt VGCUT}$, and $m_q, m_g$ being the quark and gluon effective masses, and {\tt VQCUT,VGCUT} the minimum virtuality parameters. Similar cuts are applied for initial state shower.

\begin{figure}[h!tb]
\begin{center} 
\includegraphics[width=0.45\textwidth]{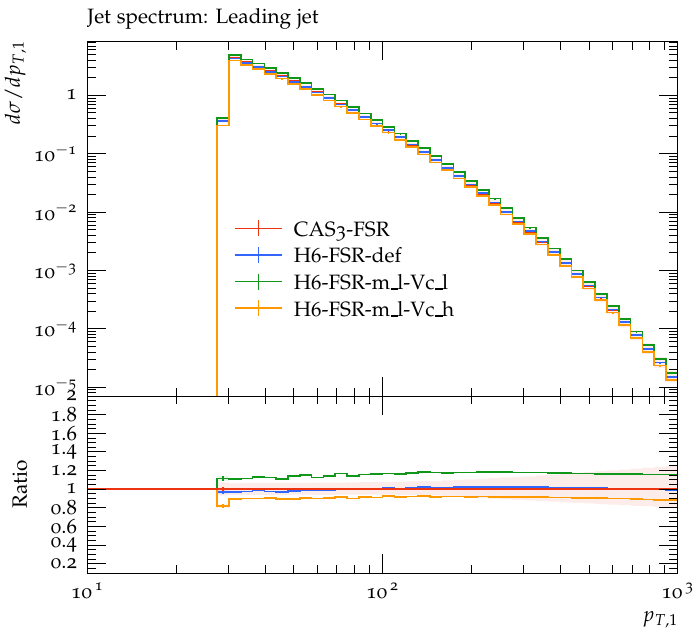} 
\includegraphics[width=0.45\textwidth]{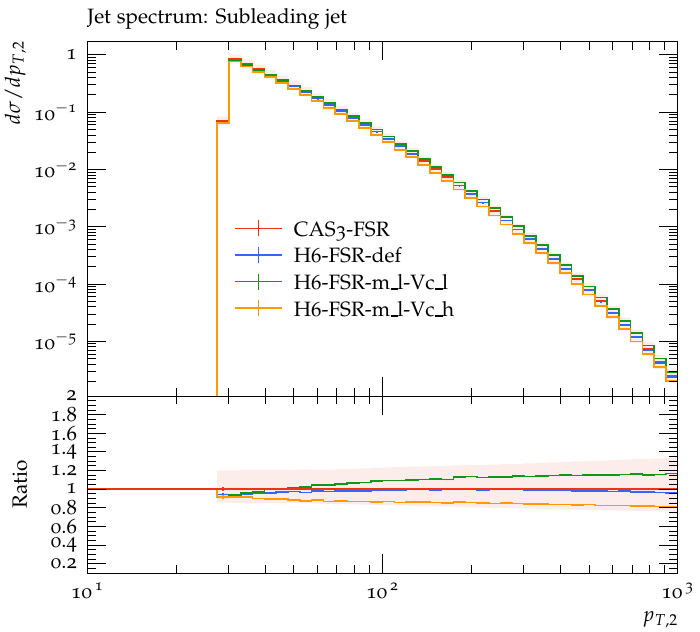} 
  \caption{\small Comparison of predictions obtained with \protect\mcatnlo+CAS3 and \protect\mcatnlo+H6 for \Zjet\ obtained with \mcatnlo .
  Shown are predictions using only final state parton shower. The band of \protect\mcatnlo+CAS3 shows the uncertainties obtained from scale variation (as described in the text). }
\label{FSR_CAS_H6_pt}
\end{center}
\end{figure}

First we investigate final state parton showers. 
We compare distributions of the first and second jet in \Zjet\ events: the first (highest \pt ) jet is part of the lowest order process, while the second (highest \pt ) jet is the real correction and therefore subject to subtraction terms (keeping in mind that the highest \pt\ jet in the NLO calculation can also come from the  $\alpha_s^2$ real emission diagram). 
In \cascade,  the \pythia6 final state shower  is used (since the \PBM\ - method has not yet been applied for final state radiation), with the angular ordering veto condition. Since final state radiation is independent of  parton densities, a direct comparison of \mcatnlo+CAS3 and \mcatnlo+H6, using the same LHE files, while only simulating final state radiation, is possible.
In Fig.~\ref{FSR_CAS_H6_pt} we show a comparison of predictions for the transverse momentum of the first two highest \pt\ jets in \Zjet\ events (using identical LHE files). 

The uncertainty coming from different parameter settings in the H6 final state parton shower is estimated by changing the light quark masses  from the default to 0.32 \GeV ({\tt Rmas = 0.32}, labelled as $m_l$) and {\tt VQCUT,VGCUT} from the default to {\tt 0.1(1.5)}, labelled as $Vc_l(Vc_h)$, respectively (the lowest values chosen are those for which H6 is still working). 

\begin{figure}[h!tb]
\begin{center} 
\includegraphics[width=0.45\textwidth]{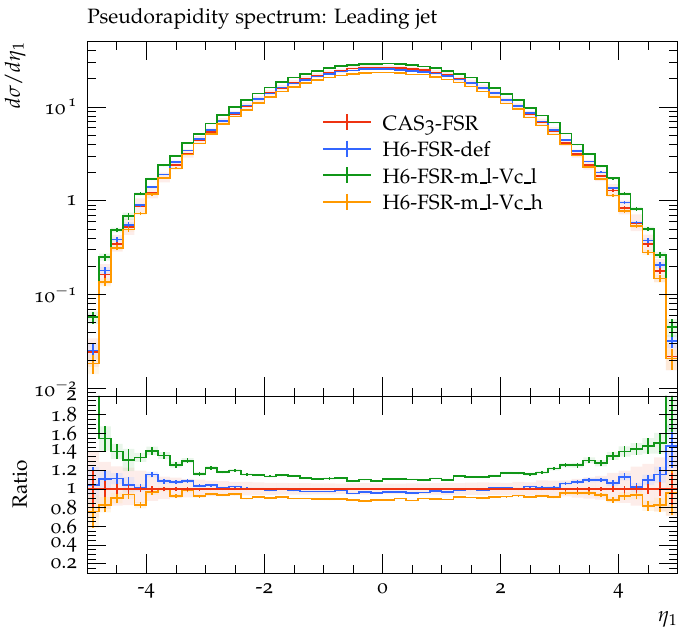} 
\includegraphics[width=0.45\textwidth]{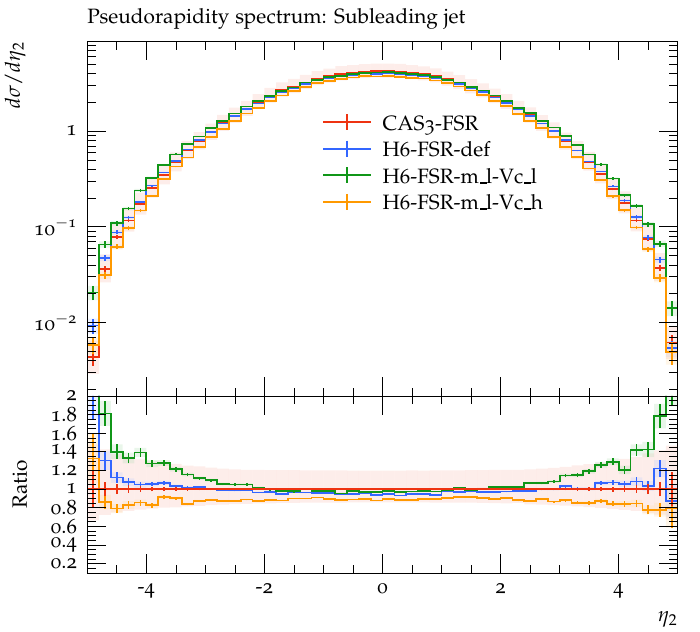} 
  \caption{\small Comparison of predictions obtained with \protect\mcatnlo+CAS3 and \protect\mcatnlo+H6 for \Zjet\ obtained with \mcatnlo .
  Shown are predictions using only final state parton shower.    The band of \protect\mcatnlo+CAS3 shows the uncertainties obtained from scale variation (as described in the text).}
\label{FSR_CAS_H6_eta}
\end{center}
\end{figure}

In Fig.~\ref{FSR_CAS_H6_eta} a comparison is shown for  the pseudorapidity $\eta$ of the first two highest \pt\ jets. Within the variation of the parameters, the prediction of \mcatnlo+CAS3 agrees well with the one of \mcatnlo+H6, justifying the application of the \pythia6 final state parton shower algorithm.

\begin{figure}[h!tb]
\begin{center} 
\includegraphics[width=0.32\textwidth]{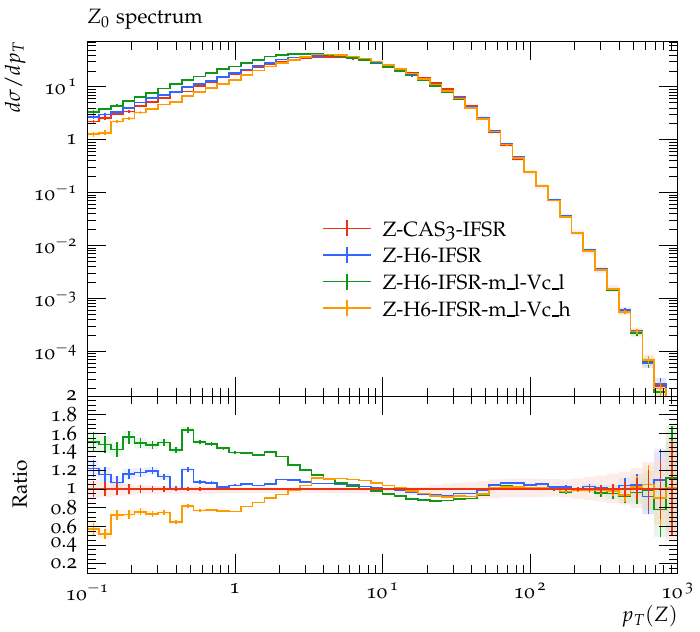} 
\includegraphics[width=0.32\textwidth]{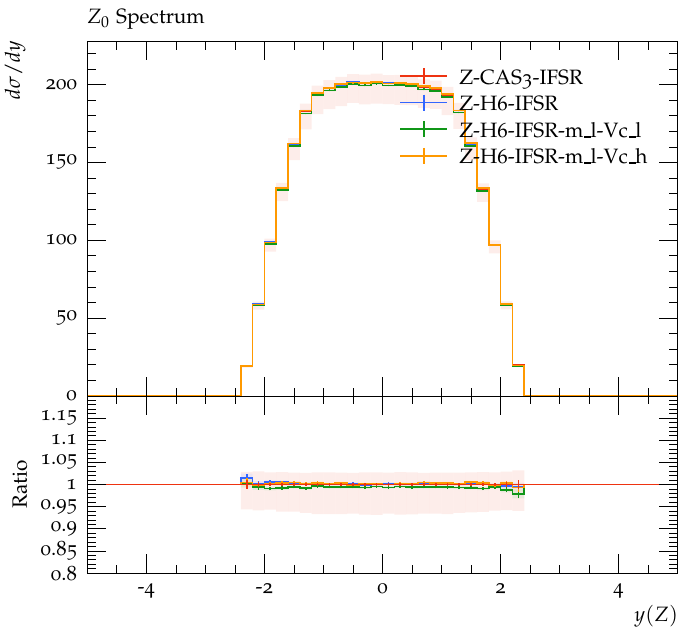}
\includegraphics[width=0.32\textwidth]{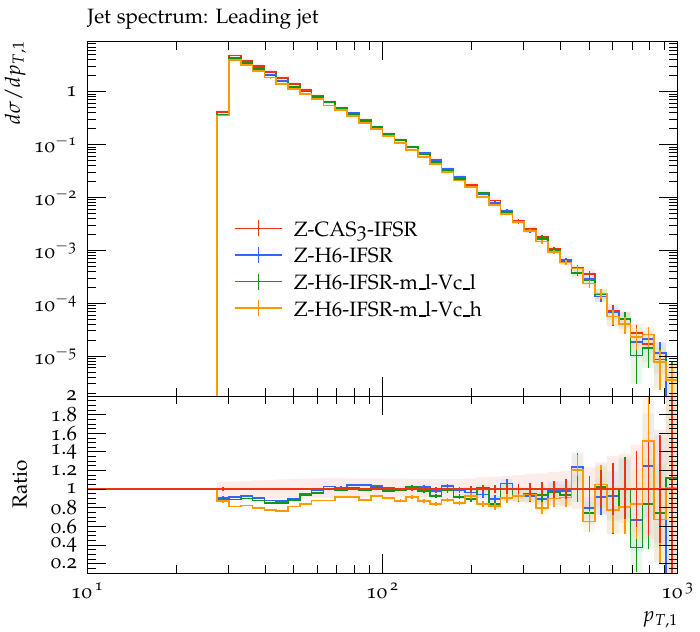} 
  \caption{\small Comparison of predictions obtained with \protect\mcatnlo+CAS3 and \protect\mcatnlo+H6 for \PZ\ production obtained with \mcatnlo .
  Shown are predictions using initial state parton shower.  The band of \protect\mcatnlo+CAS3 shows the uncertainties obtained from scale variation (as described in the text).}
\label{ISR_CAS_H6}
\end{center}
\end{figure}

Next we investigate the contribution of \PBM\ - TMD PDFs and the \PBM\ - TMD parton shower in the initial state and compare the predictions with the ones from H6.  We study  \PZ\ production generated by  \MCatNLO\   which is essentially driven by initial state radiation.  
In Fig.~\ref{ISR_CAS_H6} we show the transverse momentum of the \PZ boson, its rapidity distribution and the transverse momentum of the first reconstructed jet with $\pt > 30$ \GeV\ and $|\eta| < 5 $. Here  the rapidity $y$ of the  \PZ\ boson is used,  since it is related to the momentum fractions of the initial partons (instead of the pseudorapidity $\eta$ which is used for jets as it is related to the scattering angle $\theta$).
We show a comparison of \mcatnlo+CAS3 and \mcatnlo+H6 predictions (including the same parameter variations for H6 as for the final state shower). In the region of low transverse momentum of the \PZ\ boson one can clearly see the sensitivity to the parameter choice in H6.
While at low \pt\ the parton shower matters, and the CAS3 prediction lies in between the one from H6 with parameter variation, we observe good matching of the parton shower to the real emission at higher transverse momentum. The rapidity distribution obtained from CAS3 also lies within the one predicted by H6 with parameter variation. The \pt\ distribution of the first jet  also agrees well within the band given by the uncertainties.

\begin{figure}[h!tb]
\begin{center} 
\includegraphics[width=0.45\textwidth]{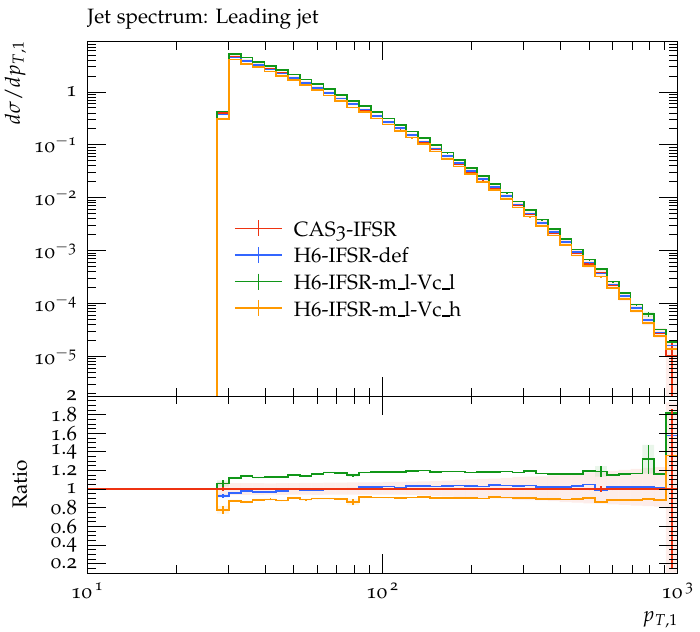} 
\includegraphics[width=0.45\textwidth]{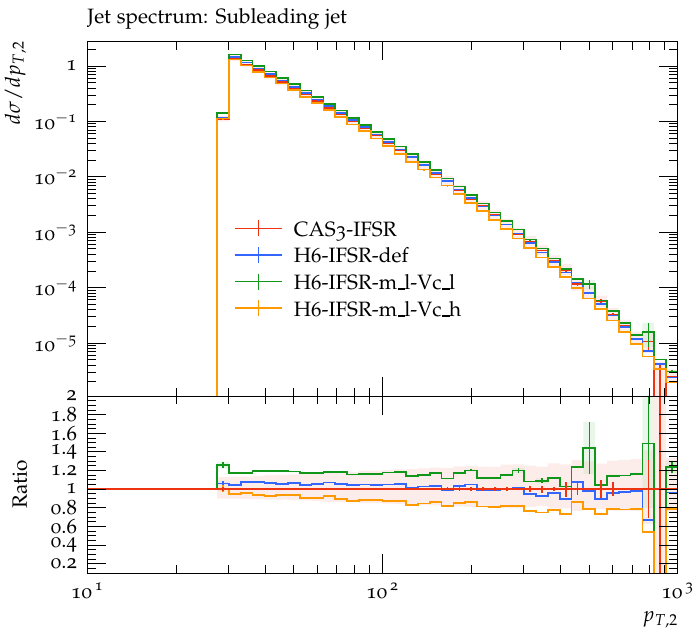} 
  \caption{\small Comparison of predictions obtained with \protect\mcatnlo+CAS3 and \protect\mcatnlo+H6 for \Zjet\ obtained with \mcatnlo .
  Shown are predictions using initial and final state parton shower.  The band of \protect\mcatnlo+CAS3 shows the uncertainties obtained from scale variation (as described in the text).}
\label{IFSR_CAS_H6_pt}
\end{center}
\end{figure}

\begin{figure}[h!tb]
\begin{center} 
\includegraphics[width=0.45\textwidth]{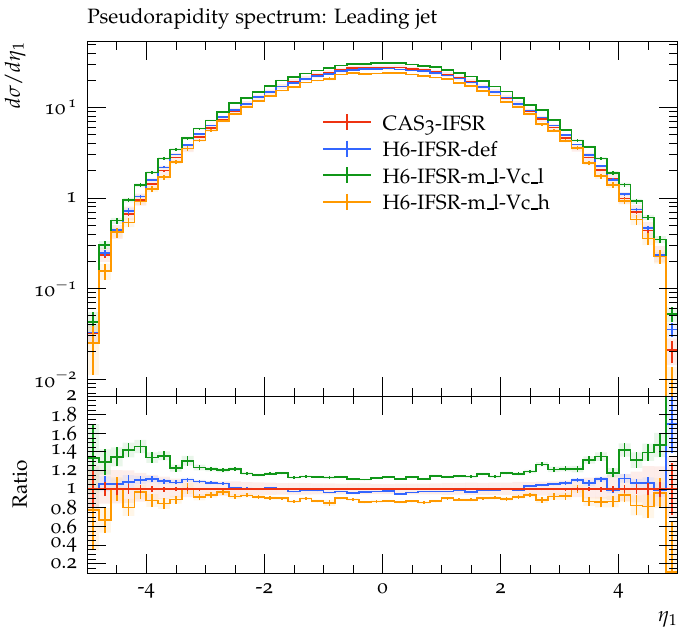} 
\includegraphics[width=0.45\textwidth]{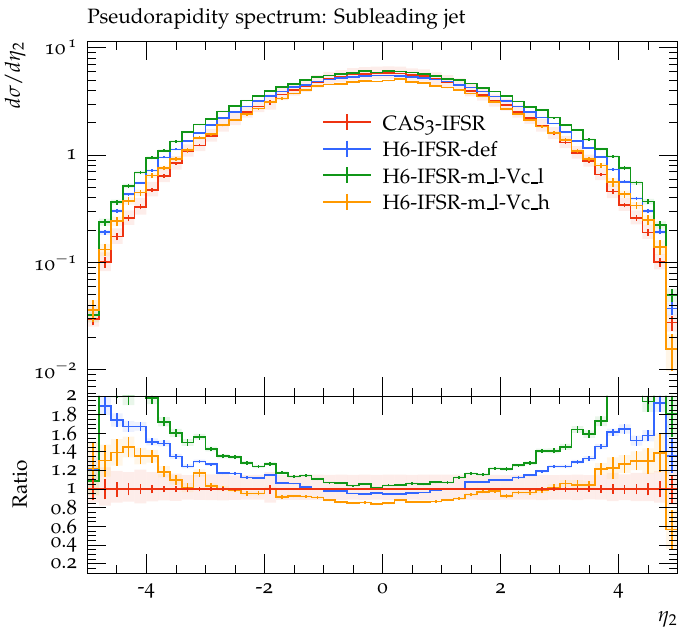} 
  \caption{\small Comparison of predictions obtained with \protect\mcatnlo+CAS3 and \protect\mcatnlo+H6 for \Zjet\ obtained with \mcatnlo .
  Shown are predictions using initial and final state parton shower.  The band of \protect\mcatnlo+CAS3 show the uncertainties obtained from scale variation (as described in the text).}
\label{IFSR_CAS_H6_eta}
\end{center}
\end{figure}

Finally we investigate \Zjet\ events, when both initial and final state radiation is important.
In Fig.~\ref{IFSR_CAS_H6_pt} we show a comparison of \mcatnlo+CAS3 and \mcatnlo+H6 predictions (including the same parameter variations for H6 as for the final state shower) for the transverse momentum of the first two highest \pt\ jets. In Fig.~\ref{IFSR_CAS_H6_eta} the corresponding comparison is shown for the pseudorapidity distributions. The transverse momentum distributions agree well within the uncertainties coming from parameter variations, while for the $\eta$-distributions some differences in the very forward/backward regions are seen. However, one can see that a variation of {\tt VQCUT,VGCUT} has a significant effect especially in the  forward/backward region. 

In conclusion, we observe agreement between predictions obtained by  \mcatnlo+CAS3 and \mcatnlo+H6 within the band of parton shower parameter variation in H6, confirming the use of H6 subtraction terms in \mcatnlo\ together with  \PBM\ - TMD PDFs,  \PBM\ - TMD initial state parton shower, as applied in \mcatnlo+CAS3.

\clearpage

\providecommand{\href}[2]{#2}\begingroup\raggedright\endgroup

\end{document}